\documentstyle[12pt,amssymb,psfig]{article}
\newcommand{\be}{\begin{equation}}
\newcommand{\ee}{\end{equation}}
\newcommand{\pr}{\prime}

\newcommand{\p}{\partial}
\newcommand{\f}{\frac}
\newcommand{\r}{\rightarrow}

\newcommand{\bt}{\beta}

\newcommand{\z}{\zeta}
\begin{document}
\begin{titlepage}
\begin{flushright}
\begin{tabular}{l}
NTUA-102/00\\
hep-lat/0012028\\
\end{tabular}
\end{flushright}

\vskip2truecm

\begin{center}
\begin{large}
{\bf Phase Structure of the 5D Abelian Higgs Model with
Anisotropic Couplings}

\end{large}
\vskip1truecm

P.~Dimopoulos$^{(a)}$\footnote{E-mail: pdimop@central.ntua.gr},
K.~Farakos$^{(a)}$\footnote{E-mail: kfarakos@central.ntua.gr},
C.~P.~Korthals-Altes$^{(b)}$\footnote{E-mail:
altes@cpt.univ-mrs.fr}, G.~Koutsoumbas$^{(a)}\footnote{E-mail:
kutsubas@central.ntua.gr}$ and S.~Nicolis$^{(c)}\footnote{E-mail:
nicolis@celfi.phys.univ-tours.fr}
$

\vskip1truecm

{\sl $^{(a)}$ Physics Department, National Technical University\\
15780 Zografou Campus, Athens, Greece

\vskip0.5truecm

$^{(b)}$ Centre de Physique Th\'eorique, CNRS-Luminy\\ Case 907,
F-13288 Marseille Cedex 9, France

\vskip0.5truecm

$^{(c)}$ CNRS-Laboratoire de Math\'ematiques et Physique
Th\'eorique (UMR 6083)
\\ Universit\'e de Tours, Parc Grandmont, 37200 Tours,France}

\end{center}
\vskip1truecm

\begin{abstract}

We establish the phase diagram of the five-dimensional
anisotropic Abelian Higgs model by mean field techniques and
Monte Carlo simulations. The anisotropy is encoded in the gauge couplings 
as well as in the Higgs couplings. 
In addition to the usual bulk phases (confining, Coulomb and Higgs)
we find four-dimensional ``layered'' phases (3-branes) 
at weak gauge coupling, where
the layers may be in either the Coulomb
or the Higgs phase, while the transverse directions are confining. 
\end{abstract}
\end{titlepage}

\section{Introduction}
The phase structure of theories with only global symmetries turns out to
be insensitive to anisotropies  in their couplings-e.g. the Ising model
has the same critical properties on a square as well as on a rectangular
lattice. For gauge theories, however, anisotropy turns out to lead to
radical changes in the phase diagram. In four dimensions this implies
breakdown of Lorentz invariance and is therefore physically uninteresting;
for theories in higher dimensions, however, this objection no longer holds.
Since all attempts towards unification involve  theories defined in
more than four dimensions, it is of interest to explore the phase structure
of such theories in order to find four dimensional theories that are
physically interesting. 
Until now the general approach has been a $(4+n)$ dimensional space-time
with $n$ compactified dimensions; the 
vacuum is of the form $M^{1,3}\!\times\! X^n$
where $M^{1,3}$ is four-dimensional Minkowski space-time and $X^n$
is an internal compact space. Four-dimensional gravity exists as long as
the volume of the internal space is finite. 
The models with {\em non-compact} internal 
spaces usually suffer from naked singularities.
However, there are also cases where such singularities have a physical 
interpretation as is the case of delta-function singularities whose  
physical meaning invokes extended structures, such as domain walls 
and strings. This is
the case  in the Randall-Sundrum (RS) model, where the
singularities may be interpreted as a four-dimensional  domain wall, a
three-brane, embedded in a five dimensional bulk \cite{RS}. 
In that case, although the internal space may be non-compact, a
four-dimensional graviton exists and it is localized at the three-brane.
However there are interesting subtleties to this picture that have recently
been the topic of intense study~\cite{porrati}.

The question one would now like to answer is, whether 
localization on the 3-brane exists for other
fields, like gauge fields, fermions and scalars.    
In searching for localized four-dimensional fields, the
equations of motion of the bulk fields which are coupled to the
background geometry are solved \cite{eombulkfields}. 
These equations have solutions which represent
localized four-dimensional massless and/or massive fields for scalars and
fermions. For gauge fields things are more complicated-cf. the recent 
paper by Kehagias and Tamvakis in ref.~\cite{eombulkfields}.

In our study we wish to explore non-perturbative features, so we define
our theory on the lattice and use Monte Carlo simulations to study the phase 
diagram of gauge fields coupled to scalars in five dimensions.
In order to acquire an 
intuitive understanding we shall also use mean field theory.
Both have already shed considerable light for the case of
pure gauge theories~\cite{funielsen,lowe,hulsaltesnic,difarkekuts} as well as
when 
fermions are included~\cite{hulsaltesnic}. 
A new phase has been discovered in the pure 
U(1) gauge theory in 5 dimensions, where four dimensional ``layers'' in the
{\em Coulomb} phase are separated from each other by a confining force. 
In particular, the transition from the five-dimensional confining phase to the 
layered phase turns out to
be of {\em second} order~\cite{difarkekuts,arjan}, implying the
existence of new continuum theories (for suggestions in these directions
cf.~\cite{ambjorn}). 
These features survive when fermions are introduced as 
well~\cite{hulsaltesnic,arjan}.

In what follows we shall present evidence for the existence of
a new layered phase: the layers may be in the
{\em Higgs} phase, separated from each other by a confining force.
A preliminary, three-dimensional, version of this model has already
appeared in \cite{dfk} in connection with a possible application
to condensed matter 2-D systems in strong magnetic fields.

The plan of the paper is as follows:
In section \ref{formul} we write down the model and we
recall the phase diagram for the pure U(1) gauge theory.  
In section~\ref{mfa} we use mean field theory to map out the phase
diagram and study the order of the transitions between the different phases
when scalar matter is included; 
in section~\ref{mc} we use Monte Carlo simulations to go beyond the
limitations of the mean field approximation and to characterize
 the possible phase transitions more precisely.

\section{Formulation of the model} \label{formul}

The model under study is the Abelian Higgs model in the five-dimensional 
space. Direction $\hat{5}$ will be singled 
out by couplings that will differ from the corresponding ones
in the remaining four directions. 

We proceed with writing down the lattice action of the model.

$$S=
 \beta \sum_x\sum_{1 \le \mu<\nu \le 4}(1-\cos F_{\mu \nu}(x))
+{\beta}'\sum_x\sum_{1 \le \mu \le 4}(1-\cos F_{\mu 5}(x))
$$
$$
+\bt_{h}\sum _{x} {\rm Re} [4 \varphi^*(x)\varphi (x)
- \sum_{1 \le \mu \le 4} \varphi^*(x)U_{\hat \mu}(x) \varphi (x+\hat \mu)]
$$
$$
+{\bt_h}'\sum _{x} {\rm Re} [\varphi^*(x)\varphi (x)
- \varphi^*(x)U_{\hat 5}(x) \varphi (x+\hat 5)]
$$
\be
+\sum _{x}[(1-2\beta_{R}-4 \bt_h-{\bt_h}')\varphi^*(x)\varphi (x)
+\bt_{R}(\varphi ^*(x)\varphi (x))^2],
\label{compactaction}
\ee
where
$$
F_{\mu \nu}(x)=A_\mu(x)+A_\nu(x+\hat \mu)-A_\mu(x+\hat \nu)-A_\nu(x),
~~1 \le \mu<\nu \le 4,
$$
$$
F_{\mu 5}(x)=A_\mu(x)+A_5(x+\hat \mu)-A_\mu(x+\hat 5)-A_5(x)
~~1 \le \mu \le 4.
$$

We have allowed for different couplings in the various directions: the ones
pertaining to the fifth direction are primed to distinguish them from
the ``space-like" couplings. The fifth direction will also be called
``transverse" in the sequel.

The link variables 
$U_{\hat \mu}(x)$ are defined as $e^{\mathrm{i} a_S {\overline A_S(x)}}$ or 
$e^{\mathrm{i} a_T {\overline A_T(x)}}$ respectively, where 
${\overline A_S}(x),~~{\overline A_T}(x)$ are the continuum fields and
$a_S,~a_T$ are the lattice spacings in the space-like and
the transverse-like dimensions respectively. 
The lattice fields are $$A_S(x) \equiv a_S {\overline A_S}(x),~~
A_T(x) \equiv a_T {\overline A_T}(x).$$ 
In addition, the scalar fields are also written in the polar form 
$\varphi(x) = \rho(x) e^{i \omega(x)}.$ The order parameters that we will use are 
the following:

\be
{\rm Space-like~Plaquette:~~~} P_S \equiv \left\langle\f{1}{6 N^5} \sum_x 
\sum_{1 \le \mu<\nu \le 4} \cos F_{\mu \nu}(x)\right\rangle
\ee
\be
{\rm Transverse-like~Plaquette:~~~} P_T \equiv \left\langle\f{1}{4 N^5} 
\sum_x \sum_{1 \le \mu \le 4} \cos F_{\mu 5}(x)\right\rangle
\ee
\be
{\rm Space-like~Link:~~~} L_S \equiv \left\langle\f{1}{4 N^5} \sum_x 
\sum_{1 \le \mu \le 4} \cos(\omega(x+\hat \mu) +A_{\hat \mu}(x)-\omega(x))
\right\rangle
\ee
\be
{\rm Transverse-like~Link:~~~} L_T \equiv \left\langle\f{1}{N^5} \sum_x 
\cos(\omega(x+\hat 5) +A_{\hat 5}(x)-\omega(x))\right\rangle
\ee
\be
{\rm Higgs~field~measure~squared:~~~} \rho^2 \equiv \f{1}{N^5} \sum_x \rho^2(x)
\ee
In the above equations $N$ is the linear dimension of a symmetric $N^5$ 
lattice.

The na\"{\i}ve continuum limit of the lattice action (\ref{compactaction}) 
may be obtained as follows (where an overbar is used for the
continuum fields):
$$
\varphi={\overline \varphi} \sqrt{ \f{2 a_S^2 a_T}{\bt_{h}} }, 
$$
$$
~~ A_\mu = a_S {\overline A_\mu}, ~~1 \le \mu \le 4, 
$$
$$
~~A_5 = a_T {\overline A_5}.
$$
Then the transverse-like field strength $$F_{\mu5} \equiv A_\mu(x)+
A_5(x+\hat \mu)-A_\mu(x+\hat 5)-A_5(x)~~(1 \le \mu \le 4)$$ 
goes over to: 
$$-a_S [a_T \p_5 {\overline A_\mu}(x)]+a_T [a_S \p_\mu {\overline A_5}(x)] = 
a_S a_T (\p_\mu {\overline A_5}-\p_5 {\overline A_\mu}).$$
Thus $$F_{\mu5}^2 \r a_S^2 a_T^2 {\overline F_{\mu5}^2},
~~1 \le \mu \le 4~~(F_{\mu 5} \equiv \p_\mu {\overline A_5}
-\p_5 {\overline A_\mu}).$$
The space-like field strength is treated in a very similar way 
with the result:
$$F_{\mu\nu}^2 \r a_S^4 {\overline F_{\mu\nu}^2},~~
{\overline F_{\mu\nu}} \equiv \p_\mu {\overline A_\nu}
-\p_\nu {\overline A_\mu},~~1 \le \mu < \nu \le 4.$$

This means that the transverse-like part of the pure gauge action 
is rewritten in the form: 
$$\f{1}{2} \f{{\bt}' a_T}{a_S^2} \sum a_S^4 a_T 
[\sum_{1 \le \mu \le 4} {\overline F}_{\mu 5}^{2}] 
\rightarrow \f{1}{2} \f{{\bt}' a_T}{a_S^2} \int d^5 x [\sum_{1 \le \mu \le 4} {\overline F}_{\mu 5}^{2}].$$
On the other hand the space--like part is:
$$\f{1}{2} \f{\bt}{a_T} \sum a_S^4 a_T
[\sum_{1 \le \mu <\nu \le 4} {\overline F}_{\mu \nu}^{2}] 
\rightarrow \f{1}{2} \f{\bt}{a_T} \int d^5 x [\sum_{1 \le \mu < \nu \le 4} 
{\overline F}_{\mu \nu}^{2}].$$
If we define
\be
\bt_{g} \equiv \f{a_T}{g_{S}^2},
~~{\bt}' \equiv \f{a_S^2}{g_{T}^{2} a_{T}},
\ee
the resulting continuum action reads:
$$\f{1}{2} \int d^5 x \left [ \f{1}{g_S^2} \sum_{1 \le \mu < \nu \le 4} {\overline F_{\mu\nu}}^2 +\frac{1}{g_{T}^2} \sum_{1 \le \mu \le 4} {\overline F_{\mu5}}^2 \right ]$$
Defining $\gamma_{g} \equiv (\frac{{\beta}' }{\beta})^{1/2}$ and using
the definitions of $\beta$, ${\beta}'$ we find that 
$$\gamma_{g}= \frac{g_S}{g_T}\frac{a_S}{a_T}.$$ 
We denote by $\xi$ the important ratio 
$\frac{a_S}{a_T}$ of the two lattice spacings (the {\it correlation anisotropy 
parameter}) and finally derive the relation:
$$\gamma_{g}=\sqrt{ \f{{\beta}'}{\bt} } 
= \frac{g_S}{g_T} \xi.$$
Along the same lines, one may rewrite the scalar sector of the action in the 
form:
\be
\int d^5 x [\sum_{1 \le \mu \le 4} |D_\mu {\overline \varphi}|^2 
+\f{\gamma_\varphi^2}{\xi^2} |D_5 {\overline \varphi}|^2+
m^2 {\overline \varphi}^* {\overline \varphi} 
+ \lambda ({\overline \varphi}^* {\overline \varphi})^2 ],
\label{contscal}
\ee
where $D_\mu \equiv \p_\mu-i {\overline A_\mu},~~1 \le \mu \le 5.$

We have used the notations: 
$$ \gamma_\varphi \equiv \sqrt{ \f{{\beta_h}'}{\beta_{h}} },$$
$$ m^2 a{_S}^{2} \equiv \frac{2}{\beta_{h}}(1-2\beta_{R}
-4\beta_{h}-{\beta_h}'),~~\f{\lambda}{a_S}=\f{4\beta_R}{ \beta_{h}^2 \xi}.$$

In this paper we don't touch the problem of quantum corrections to the
dependence of the space-like and transverse-like couplings on the 
lattice spacings $a_S~{\rm and}~a_T,$ but we just consider the tree 
level relations derived above.
Moreover, we choose a common value for the gauge coupling constants:
$g_S=g_T \equiv g,$ (so that $\gamma_g=\xi,$)  
and assume that all the covariant derivatives 
in equation (\ref{contscal}) have the same factor in front: $\gamma_\phi=\xi,$
in accordance with the tree-level relations. Notice that the quantum corrections
give in general a $\xi$-dependence in the effective 
$g_S,~g_T$ couplings \cite{karsch}.
We note that in some runs we will vary the quantities 
$\beta,~{\beta}',~\beta_{h},~{\beta_h}'$ in such a way that 
we have $$ \gamma_\varphi=\gamma_g \equiv \zeta,~\bt^\prime= 
\bt \z^2,~\bt_h^\prime= \bt_h \z^2,\label{scal} $$
while the parameter $\beta_R$ is found
from the equation $\beta_R = \f{x \beta_{h}^2}{4 \beta},$
using the fixed value $x=2$ for the parameter
$x \equiv \f{\lambda}{g^2}.$
It should be noted that our choice of parameters does not necessarily lead 
to an isotropic continuum theory, unless $\xi=1.$

Let us now recall the salient features of the  pure U(1) phase diagram
with anisotropic couplings (more details may be found in 
\cite{funielsen,hulsaltesnic,difarkekuts}). 
For large values for $\bt$ and $\bt^\pr,$
the model lies in a Coulomb phase in five dimensions. There is a
Coulomb force between two test charges in this phase. Now consider what
will happen when one keeps $\bt$ constant, but lets $\bt^\pr$ take smaller
and smaller values. Nothing will change in the four directions
that have to do with $\bt,$ so the force will still be Coulomb-like;
however, the force between the test charges in the fifth
direction will increase and will eventually become confining when
$\bt^\pr$ becomes small enough. It is well known that the potential between
heavy test charges is closely connected with the Wilson loops. According to
the above description, the Wilson loops behave as follows:
\begin{enumerate}
\item $W_{\mu \nu}(L_1,L_2) \approx \exp(-\sigma L_1 L_2) $ (strong coupling)
for small values of $\bt$ and $\bt^\pr.$
\item $W_{\mu \nu}(L_1,L_2) \approx \exp(-\tau (L_1+L_2)) $
(Coulomb phase, $1 \le \mu, \nu \le 5),$ for $\bt>1$ and $\bt^\pr>0.4.$ 
\item $W_{\mu \nu}(L_1,L_2) \approx \exp(-\tau (L_1+L_2)),~~ 
W_{\mu 5}(L_1,L_2) \approx \exp(-\sigma^\pr L_1 L_2) $ 
(layered phase, $1 \le \mu, \nu \le 4,)$ for $\bt>1$ and $\bt^\pr<0.4.$
\end{enumerate}
The quantities $\sigma,~\tau,~\sigma^\pr$ are positive constants.
Let us remark here that there is no layered phase with the roles of $\bt$
and $\bt^\pr$ reversed, since the two parameters enter in a quite different
way in the action. The layered phase is due to the simultaneous
existence of Coulomb forces in the space-like directions and confining
forces in the fifth direction.

\section{Mean Field Approach}\label{mfa}

Our starting point is the action (\ref{compactaction}). 
We shall fix the gauge by imposing $U_{\hat 4}(x)=I$ and use the 
translation-invariant {\em
Ansatz}~\cite{funielsen,lowe,hulsaltesnic},
$U_{\hat \mu}(x)=v,\,1 \le \mu \le 3$; $U_{\hat 5}(x)=v'$. We also
introduce the variables for the Higgs field,
\begin{equation}
\label{Higgs}
\phi(x)=\rho(x)v_{\chi}(x)
\end{equation}
and have also assumed a translationally invariant {\em Ansatz},
$\rho=\rho(x)$, $v_{\chi}=v_{\chi}(x)$. 
The free energy, which should be minimized to get the mean field
solution, reads:
\begin{equation}
\label{fren}
\begin{array}{ll}
F=& -3\beta v_a^2\left(v_a^2+1\right)\\%%-\beta 3 v_a^4-\beta 3 v_a^2
  &-\beta'{v'_{a'}}^2\left(3v_a^2+1\right)\\
%%-\beta^\prime 3 v_a^2 v_{a'}^{\prime 2}-\beta^\prime v_{a'}^{\prime 2}
  &-(3 \beta_h v_a +\beta_h +\beta_h^\prime v_{a'}^{\prime}) \rho^2 v_\chi^2\\
  &+(1-2 \beta_R) \rho^2+\beta_R \rho^4-\frac{1}{2} \log[\rho^2]\\
  &+3 a v_a-3 \log[I_0(a)]+a^\prime v_{a'}^\prime-\log[I_0(a^\prime)]+\chi v_\chi-\log[I_0(\chi)]\\
\end{array}
\end{equation}
The parameters $a$, $a'$ and $\chi$ are conjugate to $v_a$, $v'_{a^\prime}$ and
$v_\chi$ respectively. There are three space-like plaquettes which do not 
contain $U_{\hat 4}$ and three others that do contain it; this 
explains the first line of expression (\ref{fren}).
The second line contains the expressions for the transverse-like plaquettes:
three of them contain $U_{\hat 4}$ and one does not. The third line 
refers to the three space-like links along directions $\hat 1,~\hat 2,~\hat 3,$
the one along $\hat 4,$ and the transverse-like link. The fourth line 
contains terms that do not refer to directions at all; in particular the
logarithmic last term comes from the measure of the Higgs field. Finally,
the last line has the contributions of the integration of the Haar measure:
three $a v_a-\log[I_0(a)]$ terms from the space-like links, one 
similar term with primed quantities from the transverse-like links and
one more from the angle $\chi$ of the Higgs field.

Among the results of the minimization one may single out the relations:
\begin{equation}
\label{mfeq}
\left\{
\begin{array}{c}
v_a\\ v'_{a'}\\ v_{\chi}\\
\end{array}
\right\}= u\left( \left\{
\begin{array}{c}
a\\ a'\\ \chi\\
\end{array}
\right\}\right)
\end{equation}
where
\begin{equation}
\label{mfeq1} u(z)=\left[\log
u_0(z)\right]'=\frac{u_0'(z)}{u_0(z)}=\frac{I_1(z)}{I_0(z)}
\end{equation}
and $I_k(z)$ is the modified Bessel function of order $k:$
\begin{equation}
\label{Bessel}
I_k(z)=\frac{1}{\pi}\int_{0}^{\pi}d\theta\cos(k\theta)
e^{z\cos\theta}.
\end{equation}
One may use these relations to eliminate variables $v_a$,$v'_{a^\prime}$ and
$v_\chi$ in favour of $a$, $a'$ and $\chi,$ getting an alternative form for 
the free energy.

We start our exploration using mean field theory sticking to 
the value $\bt=4.0$ for the gauge coupling within the layers, i.e. 
they're at {\em weak} gauge coupling. For {\em strong} gauge coupling 
($\bt<1.0$) preliminary results indicate a different picture, which will 
be set forth elsewhere, in order to keep the presentation clear.
$x$ has been set to 2.0 and $\bt_h$ is running. The parameters ${\bt}'$
and ${\bt_h}'$ vary according to equation (\ref{scal})
(the value of $\z$ is kept at some fixed value) 
and we choose $\beta_R=x \frac{\beta_h^2}{4\beta}.$

\begin{figure}
\centerline{\hbox{\psfig{figure=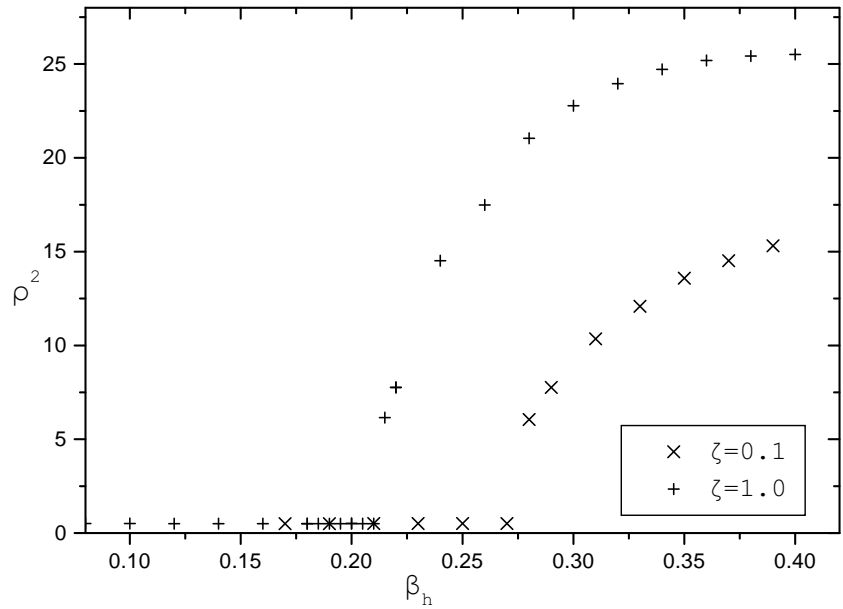,height=10cm,angle=0}}}
\caption[fm1]{$\rho^2$ versus $\beta_{h}$ for $\z=1.0$ (upper curve) 
and $\z=0.1$ (lower curve) for $\beta=4.0.$}
\label{mfig1.eps}
\end{figure}

A first set of results is given in figure \ref{mfig1.eps}.
We have calculated $\rho^2$ for two values of $\z$ by mean 
field methods: the upper curve in the figure corresponds to
$\z=1.0,$ the isotropic model, and the lower curve to a highly 
anisotropic model with $\z=0.1$ (this means that ${\bt}'=0.01 \bt,~~
{\bt_h}'=0.01 \bt_h).$ The results suggest that we have Higgs transitions 
since $\rho^2$ grows large in both curves. The fact that $\rho^2$ does 
not have the tendency to increase significantly  
as a function of $\bt_h$ after some point is not hard to
explain: it is due to the fact that $\bt_R$ is not constant, but it
increases with $\bt_h$ according to the relation $\beta_R=
x \frac{\beta_h^2}{4\beta}.$ The result is that $\rho^2$
increases as soon as the system passes to the Higgs phase and keeps
increasing for not too large values of $\bt_h.$ When $\bt_h$ grows
to even bigger values, the resulting large value of $\bt_R$ forces $\rho^2$ to
smaller values. We see that the isotropic model has the Higgs phase 
transition at $\bt_h \simeq 0.22,$ while the anisotropic model 
at $\bt_h \simeq 0.27.$

%%%%%%%%%%%%%%%%%%%%%%%%%%%%%%%%%%%%%%%%%%%%%%%%%%%%%%%%%%%%%%%%
%%%%%%%%%%%%%%%%%%%%%%%%%%%%%%%%%%%%%%%%%%%%%%%%%%%%%%%%%%%%%%%%

\begin{figure}
\centerline{\hbox{\psfig{figure=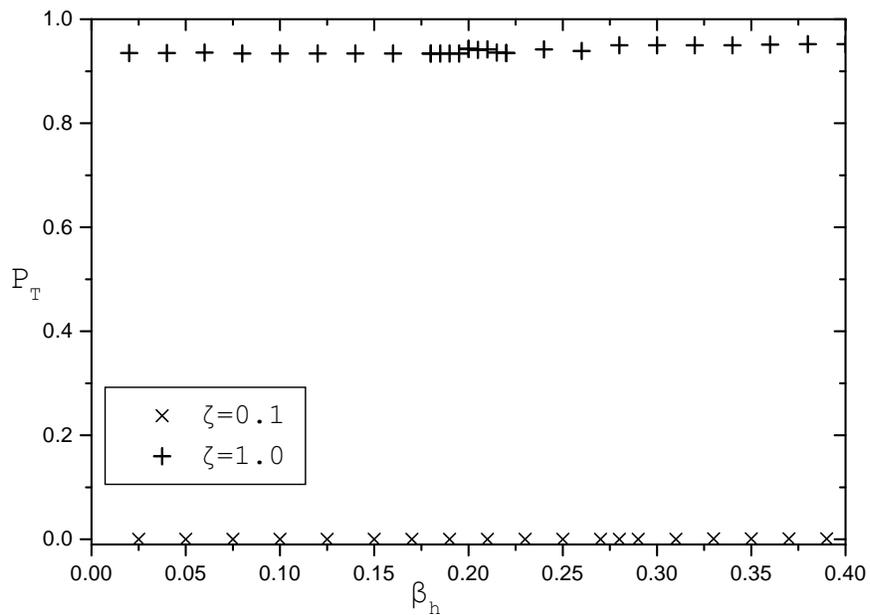,height=10cm,angle=0}}}
\caption[fm2]{Transverse-like plaquette versus $\beta_{h}$ for 
$\z=0.1$ (lower curve) and $\z=1.0$ (upper curve) 
for $\beta=4.0.$ }
\label{mfig2.eps}
\end{figure}

We should now comment on the nature of each phase;
to this end we show the behaviour of the transverse-like
plaquette, $P_T,$ in figure \ref{mfig2.eps}. 
For the isotropic model the 
transverse-like plaquette is the same as the 
space-like plaquette and it has  a (fairly constant)
big value. This means that all five directions of the model communicate 
with one another and we have a genuine five-dimensional system. 
In addition, $P_T$ has a (small) jump for 
$\bt_h=0.22$, as $\rho^2$ does in figure 
\ref{mfig1.eps} (although here it is difficult to see due to the scale 
of the present figure) and we pass from a Coulomb to a Higgs phase.
On the other hand, for the model with $\z=0.1$ the transverse-like
plaquette is very small as compared with its space-like partner. 
The picture is that the model is effectively {\em four-dimensional},
since communication between the layers is very difficult, as signalled
by the small values of the relevant quantities, such as $P_T.$
Combining the conclusions drawn from figures \ref{mfig1.eps} and  
\ref{mfig2.eps} we see that for $\z=1.0$ we have found a 
transition at $\bt_h \simeq 0.22$ separating a five-dimensional Coulomb
phase (denoted by $C_5$) from a five-dimensional Higgs phase (denoted by $H_5$).
On the other hand, for $\z=0.1,$ we have an effectively four-dimensional
system, so that the transition at $\bt_h = 0.27$ separates a four-dimensional 
Coulomb phase (denoted by $C_4$) from a four-dimensional Higgs phase 
(denoted by $H_4$). This ``world" consists of four-dimensional layers, in
which the symmetry is broken, connected with each other with confining 
forces ($\bt^\prime$ is small, so we are in the confining phase of QED
in this direction).

%%%%%%%%%%%%%%%%%%%%%%%%%%%%%%%%%%%%%%%%%%%%%%%%%%%%%%%%%%%%%%%%
%%%%%%%%%%%%%%%%%%%%%%%%%%%%%%%%%%%%%%%%%%%%%%%%%%%%%%%%%%%%%%%%

Returning to figure \ref{mfig1.eps} we observe 
that the region between its two curves contains the curves
with $0.1 < \z < 1.0.$ A good means to study the behaviour of the model 
as a function of $\z$ would be to
fix $\bt_h$ at some value and let $\z$ run from 0.0 to 1.0.
The system will go through several phases, depending on the value of
$\bt_h.$ If $\bt_h < 0.22$ the system will start from $C_4$ and at some point
it will move to $C_5$. For $0.22 \le \bt_h \le 0.27$ the system will start
with $C_4$ and will end up at $H_5$. It may pass through $C_5$ as an intermediate
step or go over directly to $H_5$.
Finally, if $\bt_h$ is fixed to some value
bigger than 0.27, the system will move from $H_4$ to $H_5$. 

\begin{figure}
\centerline{\hbox{\psfig{figure=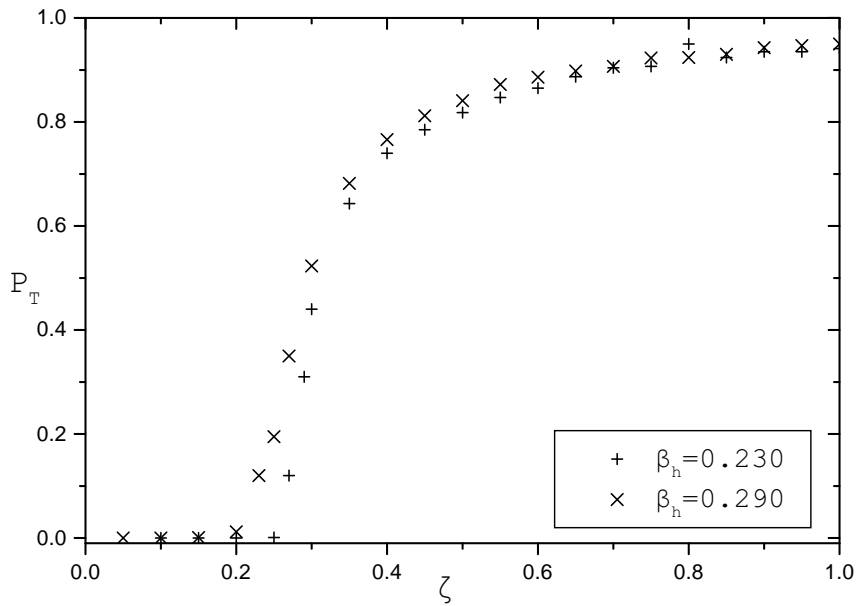,height=10cm,angle=0}}}
\caption[fm3]{Transverse-like plaquette for $\beta=4.0$
and for $\beta_{h}=0.23$ (lower curve) and
$\beta_{h}=0.29$ (upper curve).}
\label{mfig3.eps}
\end{figure}

\begin{figure}
\centerline{\hbox{\psfig{figure=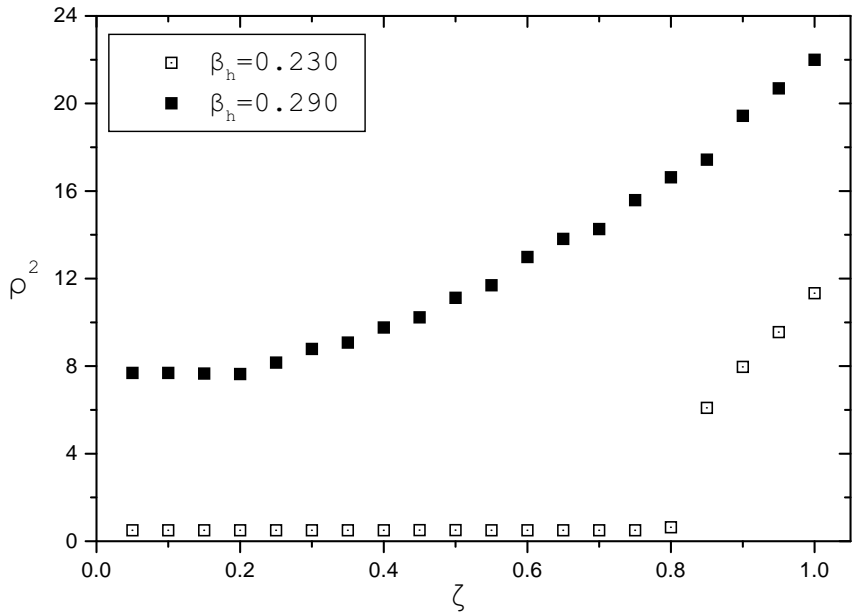,height=10cm,angle=0}}}
\caption[fm3]{$\rho^2$ for $\beta=4.0$
and the two values $\beta_{h}=0.23$ (lower curve) and
$\beta_{h}=0.29$ (upper curve).}
\label{mfig4.eps}
\end{figure}

In figure \ref{mfig3.eps} we have
plotted the transverse-like plaquette $P_T$ versus $\z.$
In the lower curve we have 
$\bt_h = 0.23,$ which lies in the interval (0.22,0.27) refered to above.
A transition is clearly visible at $\z \simeq 0.25:$ 
$P_T$ is strictly zero for $\z < 0.25$ and then
grows large, so we have a four-dimensional
system going over to a genuine five-dimensional one. A look at the 
lower curve in figure
\ref{mfig4.eps}, showing the corresponding variation of $\rho^2$
informs us that $\rho^2$ is small around this value of $\z,$ signalling
that the gauge symmetry is not broken at this transition. We conclude that 
in figure \ref{mfig3.eps}  
we see a transition from a four-dimensional Coulomb phase ($C_4$) to 
a five-dimensional Coulomb phase ($C_5$). 
Another, even stronger, phase transition, takes the system into the $H_5$
phase. We see this in figure \ref{mfig4.eps}, where $\rho^2$ 
has a discontinuity at $\z \simeq 0.8.$ 

The upper curves of figures \ref{mfig3.eps} and \ref{mfig4.eps} show
the corresponding behaviours for $\bt_h=0.29,$ where we expect a transition
from $H_4$ to $H_5$. The transverse-like plaquette in figure \ref{mfig3.eps} 
shows a smooth transition starting at $\z \simeq 0.20.$ 
$P_T$ is very small for $\z<0.20$ 
but not strictly zero as happens with the $\bt_h=0.23$ case. This is not 
possible to show in a figure, so we just mention that the typical values 
of $P_T$ are of the order of $10^{-6}$ in the $\bt_h=0.29$ case for $\z<0.20.$
This behaviour is consistent with the variation of $\rho^2,$ shown
in figure \ref{mfig4.eps}; its value varies from about 8 to about 22, both
of which characterize a Higgs phase. In addition, $\rho^2$ is constant
for the interval $0 < \z < 0.20,$ for which $P_T$ has been small.

%%%%%%%%%%%%%%%%%%%%%%%%%%%%%%%%%%%%%%%%%%%%%%%%%%%%%%%%%%%%%%%%
%%%%%%%%%%%%%%%%%%%%%%%%%%%%%%%%%%%%%%%%%%%%%%%%%%%%%%%%%%%%%%%%
%%%%%%%%%%%%%%%%%%%%%%%%%%%%%%%%%%%%%%%%%%%%%%%%%%%%%%%%%%%%%%%%
%%%%%%%%%%%%%%%%%%%%%%%%%%%%%%%%%%%%%%%%%%%%%%%%%%%%%%%%%%%%%%%%

\begin{figure}
\centerline{\hbox{\psfig{figure=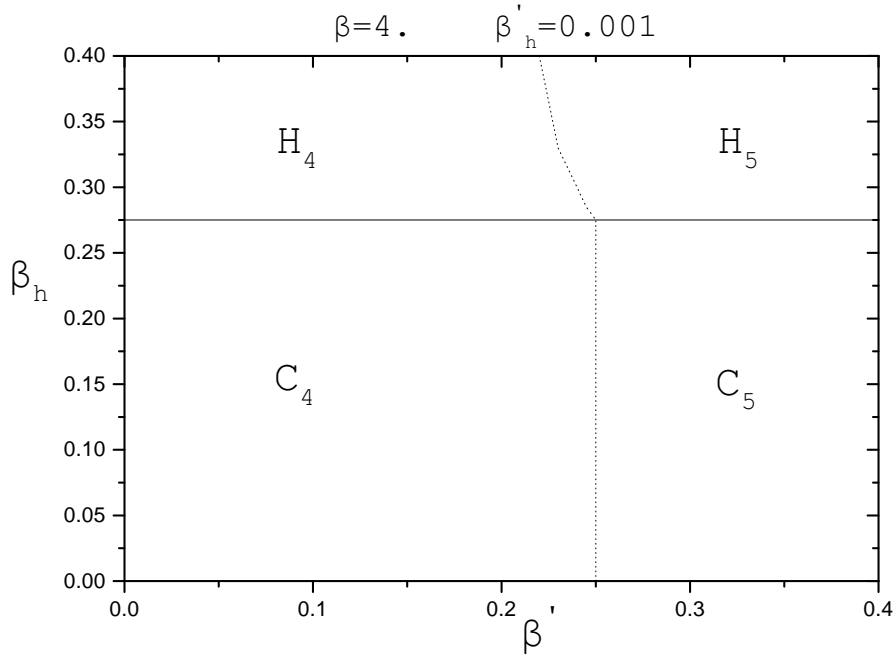,height=10cm,angle=0}}}
\caption[fm4]{Phase diagram for $\beta=4.0$ and ${\bt_h}'=0.001$.}
\label{phase.ps}
\end{figure}

Since we treat $\bt=4.0$ it would be useful to construct a phase diagram 
at $\bt=4.0$ and several values of ${\bt_h}'.$
The ${\bt_h}'$ coupling is one of the most important couplings,
since it connects the hyperplanes in the transverse direction
through the Higgs kinetic term. 
The diagram will be given in the $({\bt}'-\bt_h)$ plane. 
In figure \ref{phase.ps} we show the phase diagram for ${\bt_h}'=0.001.$ 
We observe the solid horizontal line at $\bt_h=0.27,$ which separates the
symmetric phases from the broken symmetry phases, along with the 
almost vertical dotted line, separating the four-dimensional phases from their
five-dimensional partners. We find first order phase transitions,   
one separating $C_4$ from $H_4$ and another one separating $C_5$ from $H_5.$ 
An interesting and new transition is the one from $H_4$ to $H_5$, to which 
we now turn.

%%%%%%%%%%%%%%%%%%%%%%%%%%%%%%%%%%%%%%%%%%%%%%%%%%%%%%%%%%%%%%%%
%%%%%%%%%%%%%%%%%%%%%%%%%%%%%%%%%%%%%%%%%%%%%%%%%%%%%%%%%%%%%%%%

\begin{figure}
\centerline{\hbox{\psfig{figure=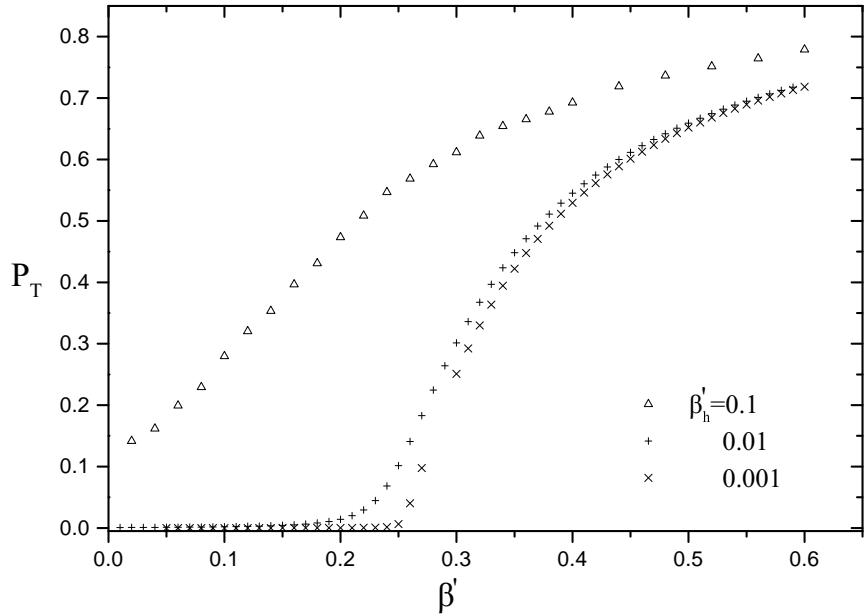,height=10cm,angle=-0}}}
\caption[fm4a]{Transverse-like plaquette versus $\bt^\prime$ 
for $\bt=4.0,~\bt_h=0.28$ and the values 0.001,0.01 and 0.10 
for the parameter $\bt_h^\prime.$}
\label{NEWfig6.eps}
\end{figure}

\begin{figure}
\centerline{\hbox{\psfig{figure=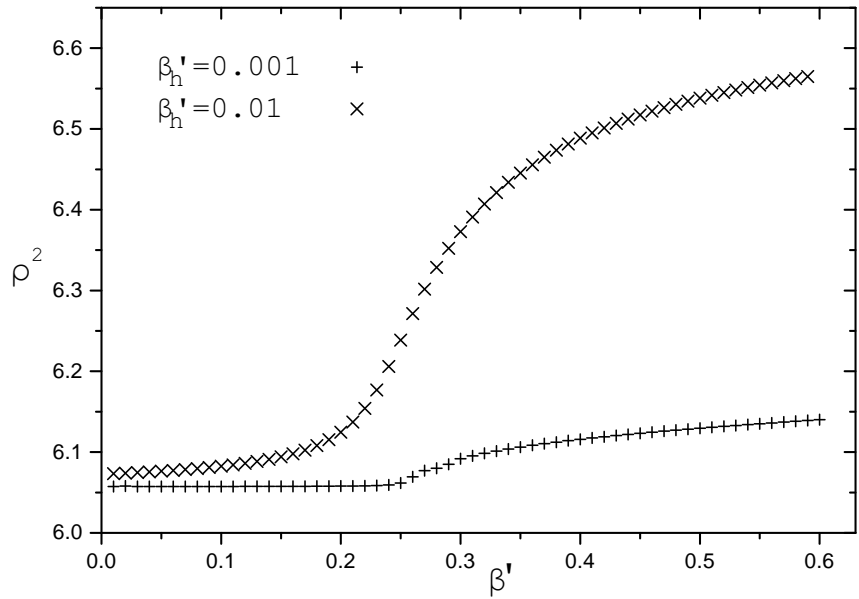,height=10cm,angle=-0}}}
\caption[fm4b]{$\rho^2$ versus $\bt^\prime$ 
for $\bt=4.0,~\bt_h=0.28$ and the values 0.001 and 0.01
for the parameter $\bt_h^\prime.$}
\label{mr2bh028.eps}
\end{figure}

Figure \ref{NEWfig6.eps} shows the variation of the transverse-like 
plaquette versus $\bt^\prime$ for $\bt=4.0,~\bt_h=0.28,~x=2$ and 
three values for $\bt_h^\prime.$ For $\bt_h^\prime=0.001$ it appears that 
the system stays in $H_4$ for $\bt^\prime < 0.24$ and then it moves 
very quickly to
$H_5.$ On the contrary, for $\bt_h^\prime=0.01$ the transition is smooth; 
it is extremely difficult to locate precisely the point of the transition.
The remnant of the phase transition happens for smaller values of 
$\bt^\prime,$ so if we constructed the analog of figure \ref{phase.ps}
for $\bt_h^\prime=0.01,$ the dotted line would move to the left and
would represent a higher order transition.
For $\bt_h^\prime=0.10$ there is no region in the
figure having the characteristics of $H_4;$ the system appears to be in the
$H_5$ phase for all $\bt^\prime$ couplings.
The same picture emerges from figure \ref{mr2bh028.eps}, 
which depicts the $\bt^\prime$ dependence of $\rho^2$ rather than $P_T$
(we only show the results for $\bt_h^\prime=0.001$ and $\bt_h^\prime=0.01,$
since the curve for $\bt_h^\prime=0.10$ lies at rather large values.)  
It is clear that the transition takes place at $\beta^\prime=0.24$
for $\beta_h^\prime=0.001$ and at a somewhat smaller value for 
$\beta_h^\prime=0.01.$

\begin{figure}
\centerline{\hbox{\psfig{figure=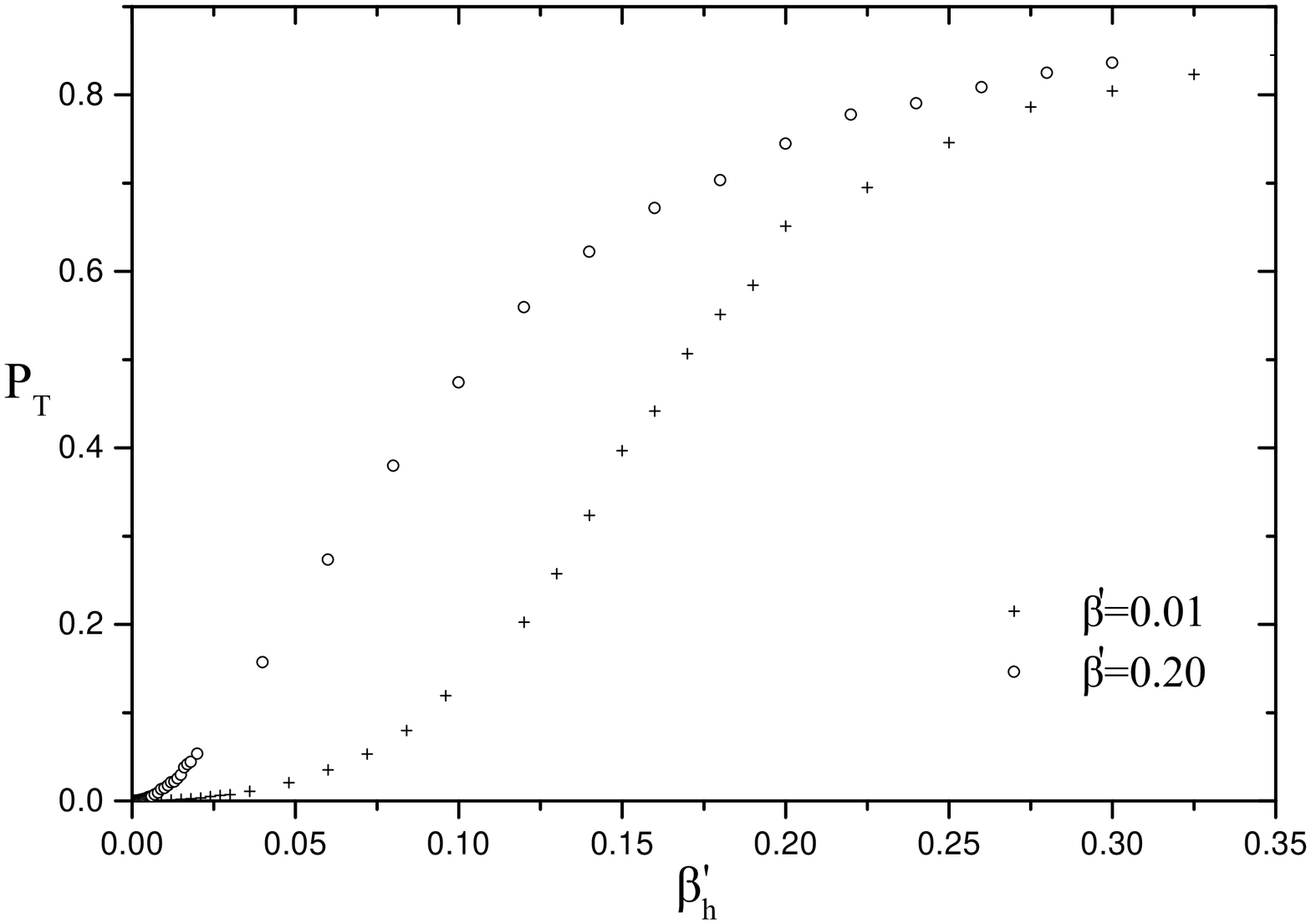,height=10cm,angle=-0}}}
\caption[fmbb]{Transverse-like plaquette versus $\bt_h^\prime$ 
for $\bt=4.0,~\bt_h=0.28.$ }
\label{fig8new.eps}
\end{figure}

We have seen that there is a remarkable difference in the behaviour 
of the system depending on the value of $\bt_h^\prime,$ so
it is natural to wonder whether there is
some phase transition when $\bt_h^\prime$ varies 
keeping the other couplings constant. 
In figure \ref{fig8new.eps} we 
display the transverse plaquette, $P_T$, vs. ${\beta_h}'$ for fixed values of 
$\beta=4.0$, $\beta_h=0.28$ and two typical values of $\beta'$ (0.2 and 0.01).
While the shape of the two curves seems to indicate that a step may appear 
for small $\beta'$, our investigation in this range shows that a 
phase transition does not occur.

%%%%%%%%%%%%%%%%%%%%%%%%%%%%%%%%%%%%%%%%%%%%%%%%%%%%%%%%%%%%%%%%
%%%%%%%%%%%%%%%%%%%%%%%%%%%%%%%%%%%%%%%%%%%%%%%%%%%%%%%%%%%%%%%%

%%%%%%%%%%%%%%%%%%%%%%%%%%%%%%%%%%%%%%%%%%%%%%%%%%%%%%%%%%%%%%%%
%%%%%%%%%%%%%%%%%%%%%%%%%%%%%%%%%%%%%%%%%%%%%%%%%%%%%%%%%%%%%%%%

%%%%%%%%%%%%%%%%%%%%%%%%%%%%%%%%%%%%%%%%%%%%%%%%%%%%%%%%%%%%%%%%
%%%%%%%%%%%%%%%%%%%%%%%%%%%%%%%%%%%%%%%%%%%%%%%%%%%%%%%%%%%%%%%%

\section{Monte Carlo Results}\label{mc}

In this section we will reproduce and corroborate the Mean Field results
of the previous section by Monte Carlo methods. We will find the same
qualitative picture, but the critical points and the orders of the phase 
transitions will be determined more precisely.
The sequence of the measurements will parallel the corresponding ones of the
previous section.

\subsection{Fixed $\zeta$}

In this set of measurements we set $\beta=4.0,~~x=2,$
and let $\beta_{h}$ run. The remaining coupling 
constants depend on the value of $\z,$ that is
(we recall that $\zeta$ is the ratio: $\zeta
=\sqrt {\frac{{\beta}'}{\beta}}=
\sqrt{\frac{{\beta_h}'}{\beta_{h}}}$):
\be
{\beta_h}'= \zeta^{2} \beta_{h},~~{\beta}'= \zeta^{2} \beta,~~
\beta_{R}=\frac{x \beta_{h}^{2}}{4 \beta}.  
\ee
Thus these measurements are a Monte Carlo realization of the mean field 
calculations depicted in figures \ref{mfig1.eps} and
\ref{mfig2.eps}, so we expect the same qualitative picture.

%%%%%%%%%%%%%%%%%%%%%%%%%%%%%%%%%%%%%%%%%%%%%%%%%%%%%%%%%%%%%%%%
\begin{figure}
\centerline{\hbox{\psfig{figure=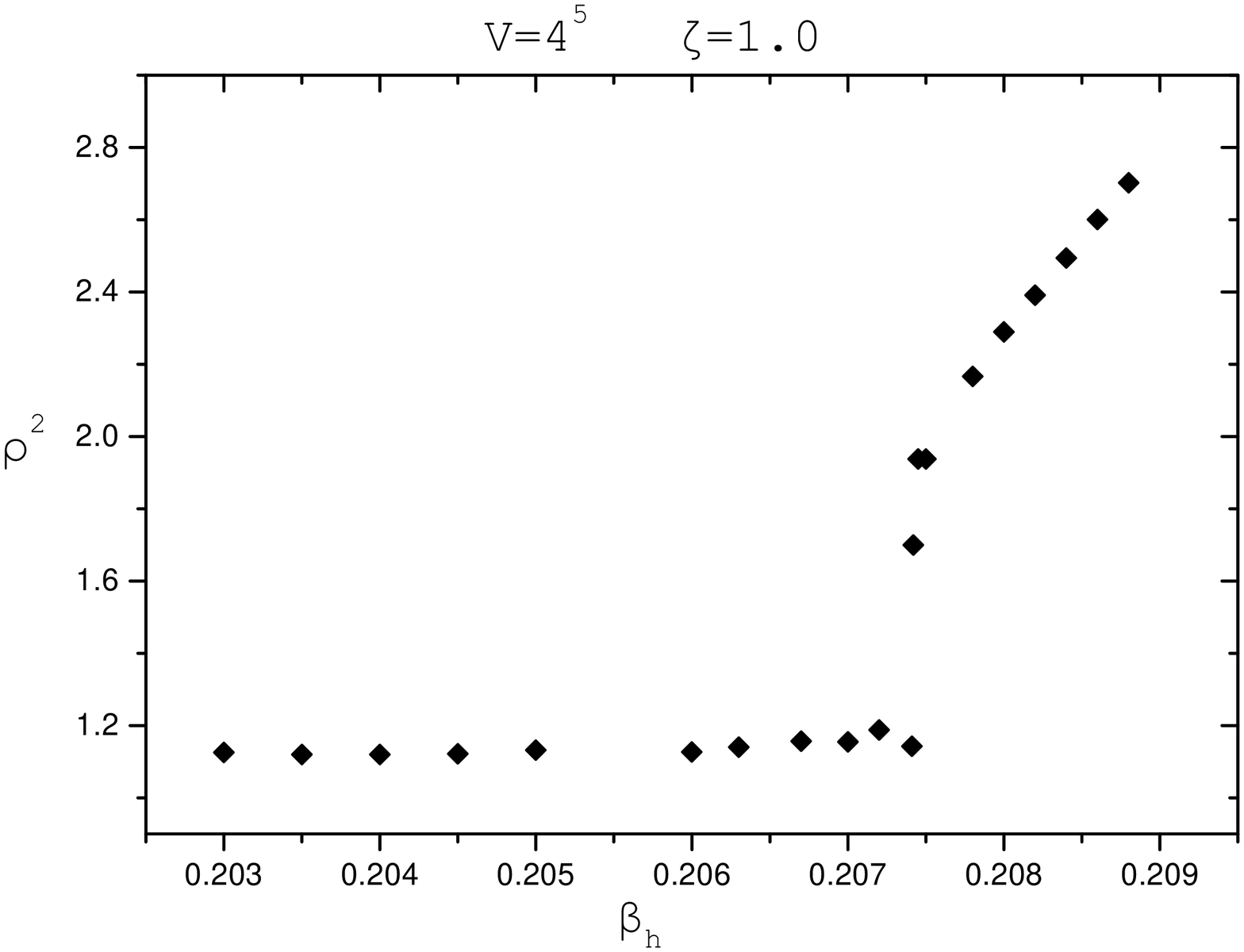,height=10cm,angle=0}}}
\caption[fn1]{$\rho^2$ versus $\beta_{h}$ for the isotropic model (
$\zeta=1.0$) for a $4^5$ lattice with $\beta=4.0.$}
\label{mr2a1.eps}
\end{figure}
%%%%%%%%%%%%%%%%%%%%%%%%%%%%%%%%%%%%%%%%%%%%%%%%%%%%%%%%%%%%%%%%

In figure \ref{mr2a1.eps} 
we show the behaviour of $\rho^2$ for the isotropic $(\z=1.0)$ model. 
The isotropic system is seen to undergo a phase transition 
at $\beta_{h} \simeq 0.2075$ from the five-dimensional Coulomb 
phase ($C_5$) to the five-dimensional Higgs phase ($H_5$). 
We note that the critical value of $\beta_{h}$ is quite close to the
mean field prediction $(\bt_h \simeq 0.22)$ for this quantity.
The Monte Carlo result in this figure shows a jump in $\rho^2,$ 
signalling a first order phase transition.

%%%%%%%%%%%%%%%%%%%%%%%%%%%%%%%%%%%%%%%%%%%%%%%%%%%%%%%%%%%%%%%%
\begin{figure}
\centerline{\hbox{\psfig{figure=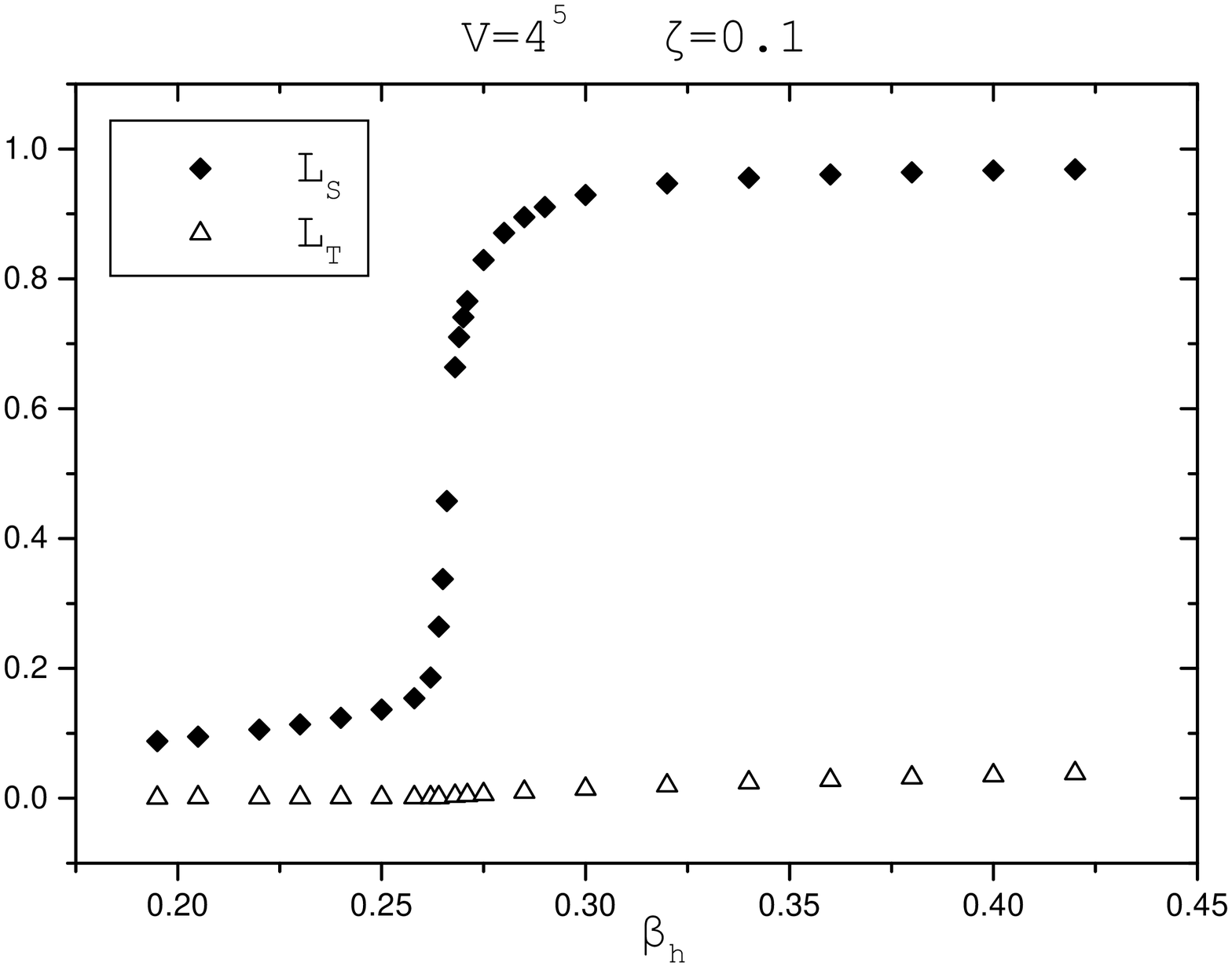,height=10cm,angle=0}}}
\caption[fn2]{Space-like and transverse-like links versus 
$\beta_{h}$ for a highly anisotropic 
model ($\zeta=0.1$) for a $4^5$ lattice with $\beta=4.0.$}
\label{mlsa01.eps}
\end{figure}
%%%%%%%%%%%%%%%%%%%%%%%%%%%%%%%%%%%%%%%%%%%%%%%%%%%%%%%%%%%%%%%%

%%%%%%%%%%%%%%%%%%%%%%%%%%%%%%%%%%%%%%%%%%%%%%%%%%%%%%%%%%%%%%%%
\begin{figure}
\centerline{\hbox{\psfig{figure=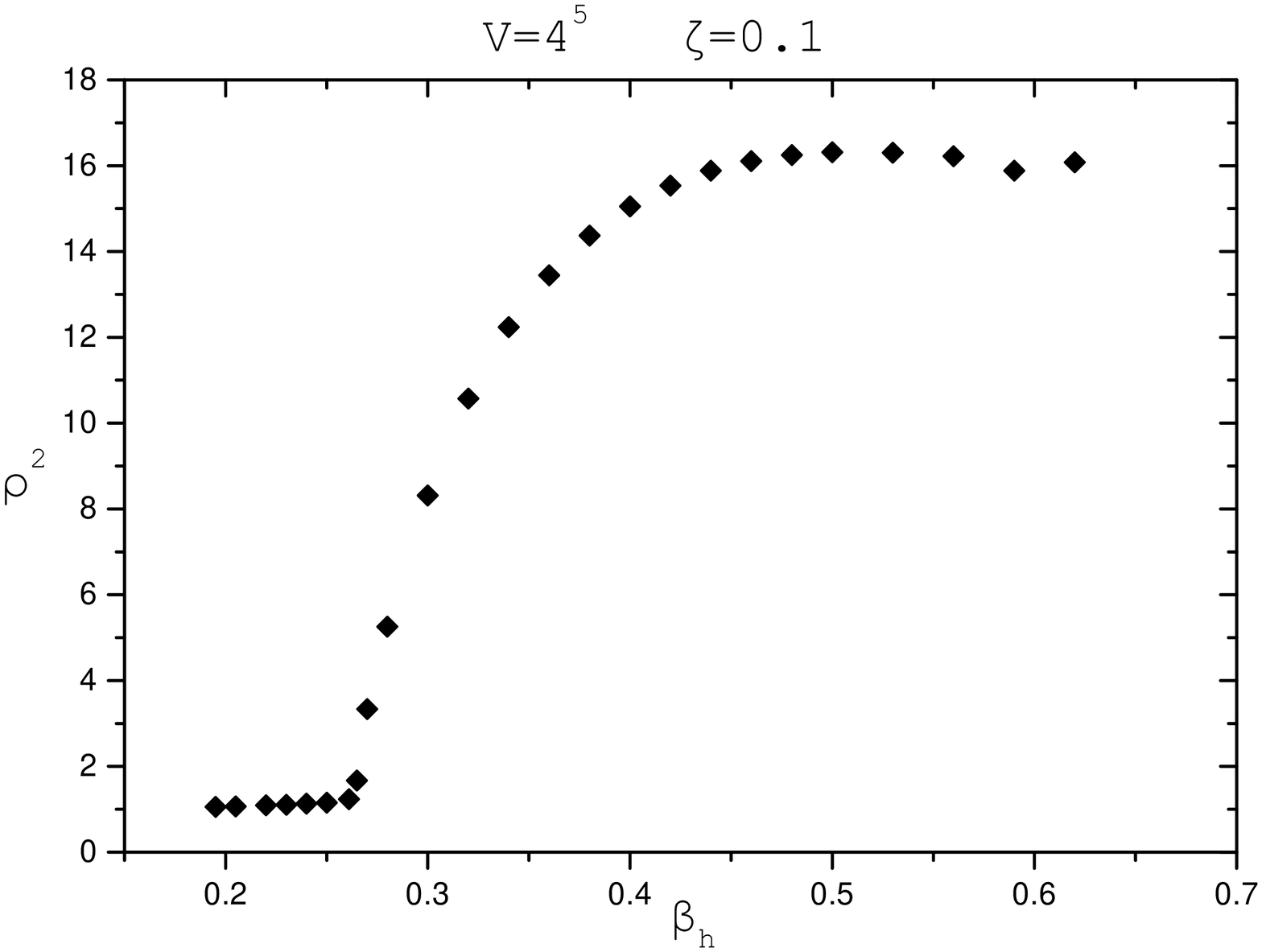,height=10cm,angle=0}}}
\caption[fn2]{$\rho^2$ versus $\beta_{h}$ for a highly anisotropic 
model ($\zeta=0.1$) for a $4^5$ lattice with $\beta=4.0.$}
\label{mr2a01.eps}
\end{figure}
%%%%%%%%%%%%%%%%%%%%%%%%%%%%%%%%%%%%%%%%%%%%%%%%%%%%%%%%%%%%%%%%

The corresponding transition for the anisotropic model ($\zeta=0.1$) 
is shown in figure \ref{mlsa01.eps}. It
takes place at $\beta_{h} \approx 0.27$  
and it seems smoother. In this case the ``transverse--like" 
coupling constants are $\zeta^2=0.01$ times smaller than their ``space--like" 
partners. This presumably means that 
the transverse--like separation $a_{T}$ of the spatial planes is  
much bigger than the spatial lattice spacing $a_{S}$. 
Support to this fact is provided by the magnitude of the quantities
$P_{S}$ and $L_{S}$ that we have measured: they turn out to be 
much bigger than $P_{T}$, $L_{T}$ for the whole 
range of $\beta_{h}$, indicating that the quantities  related to the
communication of the planes are negligible as compared against the similar 
quantities within the layers. On the other hand (compare figure 
\ref{mr2a01.eps}), $\rho^2$ is small in the  
region $\bt_h < 0.27,$ suggesting that it is a phase with unbroken symmetry. 
Based on these data and the mean field results, it seems safe to 
assume that the region 
$\beta_{h} \leq 0.27$ corresponds to a Coulomb phase, where the layers 
are decoupled: this is equivalent to the Coulomb phase of the corresponding 
{\em four--dimensional} model and will be called $C_4$ in the sequel. 
For $\beta_{h} > 0.27$ we have a Higgs phase for the anisotropic model:
$\rho^2$ is large and the quantities related to the fifth dimension 
are very small. 
This presumably means that the layers are decoupled also in this phase, 
so the picture is that we have moved effectively to the Higgs 
phase of the corresponding {\em four--dimensional} model: 
we will call this phase $H_4$. Thus in brief the anisotropic model 
moves from the 4D Coulomb phase ($C_4$) to the 4D Higgs phase ($H_4$). 
We remark that the critical parameter 
for the Higgs phase transition is of order $\frac{1}{d}$, where $d$ is the 
space--time dimension. Thus, it is not accidental that the isotropic model
has a phase transition at $\beta_{h} \simeq 0.2075$; 
this is close to the expected value, $\frac{1}{d}=0.20,$ since $d=5.$  
On the other hand, the anisotropic model is effectively 
four--dimensional, so one should expect a value approximately equal to
$\frac{1}{4}$, which is quite close to the value 0.27 given by the
simulation. 
A comparison with the mean field results is in order: we have found 
the picture predicted qualitatively by mean field theory.
In particular, we found a $C_4$-$H_4$ transition for the anisotropic model
(Monte Carlo gives it at $\bt_h \simeq 0.27$ and mean field also at
$\bt_h \simeq 0.27$). For the isotropic model the $C_5$-$H_5$ transition takes 
place at $\bt_h \simeq 0.2075$ according to the Monte Carlo and at 
$\bt_h \simeq 0.22$ according to the mean field.  

We note that the relative position of the transitions means that for 
$\beta_{h} \le 0.2075$ the systems lie in the Coulomb phase for
both models: the isotropic model in $C_5$ and the anisotropic one in $C_4$. 
For $\beta_{h} \ge 0.27$ the systems lie in their respective Higgs phases,
$H_5$ and $H_4$. Finally, for $ 0.2075 \le \beta_{h} \le 0.27$ the anisotropic 
model is in $C_4$, while the isotropic system lies in $H_5$.
An interesting question is how the various phases transform into one another 
as $\z$ varies from zero to one. We recall that the mean field approach
suggests that the transitions take place in the order 
${\rm C_4} \to {\rm C_5} \to {\rm H_5}.$ A reasonable strategy would be to fix
$\beta_{h}$ (as well as $\bt$) and vary the parameter $\z;$ this has
also been the strategy in the mean field approach.  
Figures \ref{mr2a1.eps}, \ref{mlsa01.eps}, \ref{mr2a01.eps} 
and the discussion we just made 
suggest that two possibly interesting values for $\beta_{h}$ 
would be 0.23 (to check the transition starting from the phase $C_4$ and
ending with the phase $H_5$ and determine the possible intermediate steps) 
and 0.29 (to study the transition from $H_4$ to $H_5$); we point out 
that we had chosen exactly the same values for the mean field 
calculation.  
The Monte Carlo study at the two fixed values for
$\bt_h$ is the subject of the following subsection.

\subsection{Running $\z$}

%%%%%%%%%%%%%%%%%%%%%%%%%%%%%%%%%%%%%%%%%%%%%%%%%%%%%%%%%%%%%%%%
\begin{figure}
\centerline{\hbox{\psfig{figure=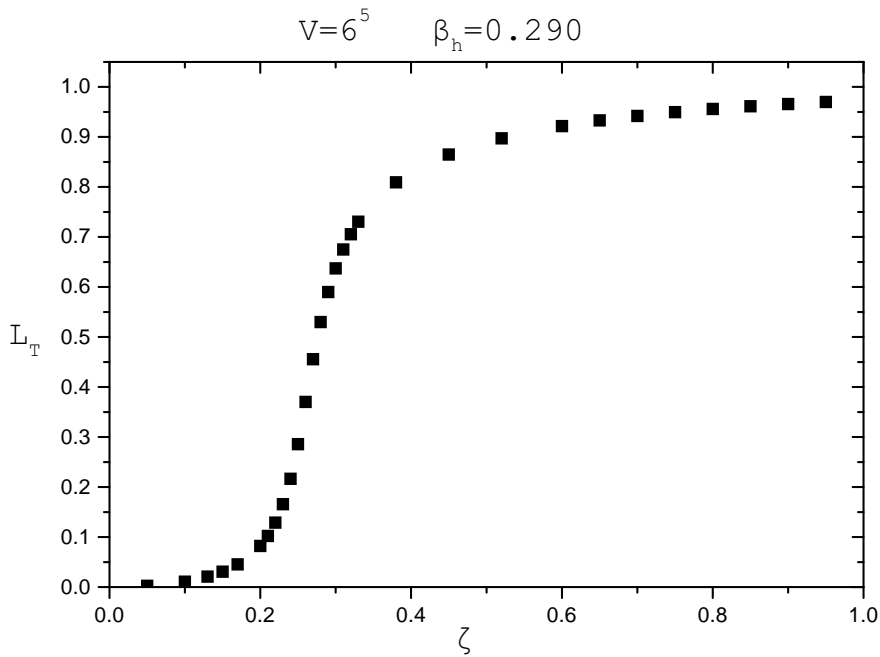,height=10cm,angle=0}}}
\caption[fn3]{Transverse-like link for a $6^5$ lattice with
$\beta=4.0,$ $\bt_h=0.29.$}
\label{mltbh0290.eps}
\end{figure}
%%%%%%%%%%%%%%%%%%%%%%%%%%%%%%%%%%%%%%%%%%%%%%%%%%%%%%%%%%%%%%%%

In figure \ref{mltbh0290.eps} 
we show the transverse-like link $L_T$ for the case with $\beta_{h} = 0.29.$ 
The parameter $\zeta$ starts from zero, where the results of the
previous subsection make us expect a Higgs phase with 
fully separated layers, that is a four-dimensional model with broken 
symmetry (denoted by $H_4$), to $\z=1.0,$ where the full five-dimensional 
Higgs phase ($H_5$) is expected on the grounds of the discussion of the previous 
subsection.
The transition from $H_4$ to $H_5$ is very smooth and takes place at about
$\z \simeq 0.25.$ Some order parameters remain almost constant
up to this value, 
and the ones that change, such as $L_T$ shown in the figure, 
do so very smoothly.

\begin{figure}
\centerline{\hbox{\psfig{figure=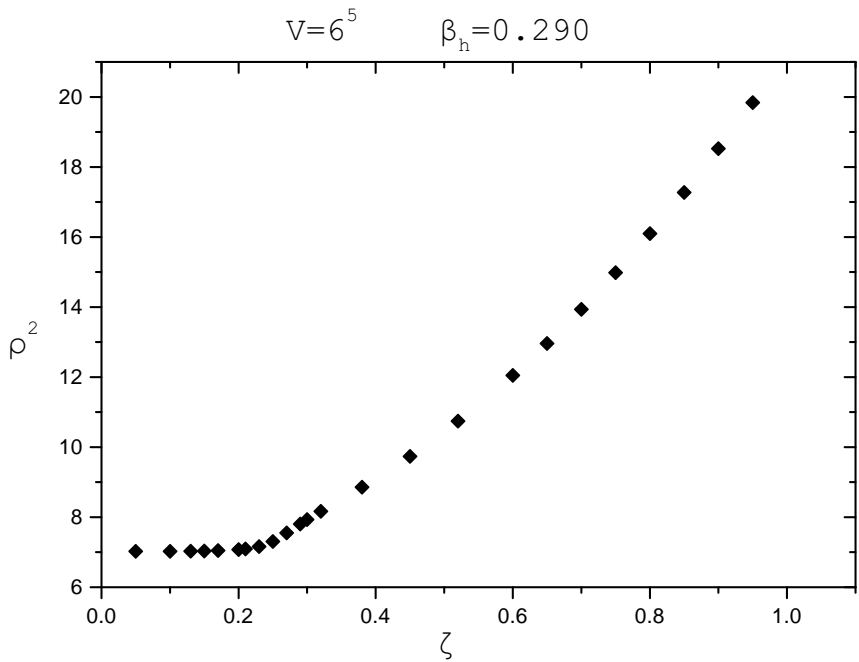,height=10cm,angle=0}}}
\caption[fn3]{$\rho^2$ for a $6^5$ lattice with
$\beta=4.0,$ $\bt_h=0.29.$}
\label{mr2bh0290.eps}
\end{figure}

Figure \ref{mr2bh0290.eps} shows another aspect of the same transition, 
namely the gradual increase of $\rho^2$ versus $\z;$ the value of this
quantity is large for the whole range of $\z,$ as would be expected for 
a transition from a Higgs phase to another Higgs phase. The important
characteristic of this quantity is the constancy of $\rho^2$
for $\z < 0.25;$ the quantities $P_S$, $L_S$ 
are also constant up to this critical value of $\z.$  

%%%%%%%%%%%%%%%%%%%%%%%%%%%%%%%%%%%%%%%%%%%%%%%%%%%%%%%%%%%%%%%%
%%%%%%%%%%%%%%%%%%%%%%%%%%%%%%%%%%%%%%%%%%%%%%%%%%%%%%%%%%%%%%%%

%%%%%%%%%%%%%%%%%%%%%%%%%%%%%%%%%%%%%%%%%%%%%%%%%%%%%%%%%%%%%%%%
\begin{figure}
\centerline{\hbox{\psfig{figure=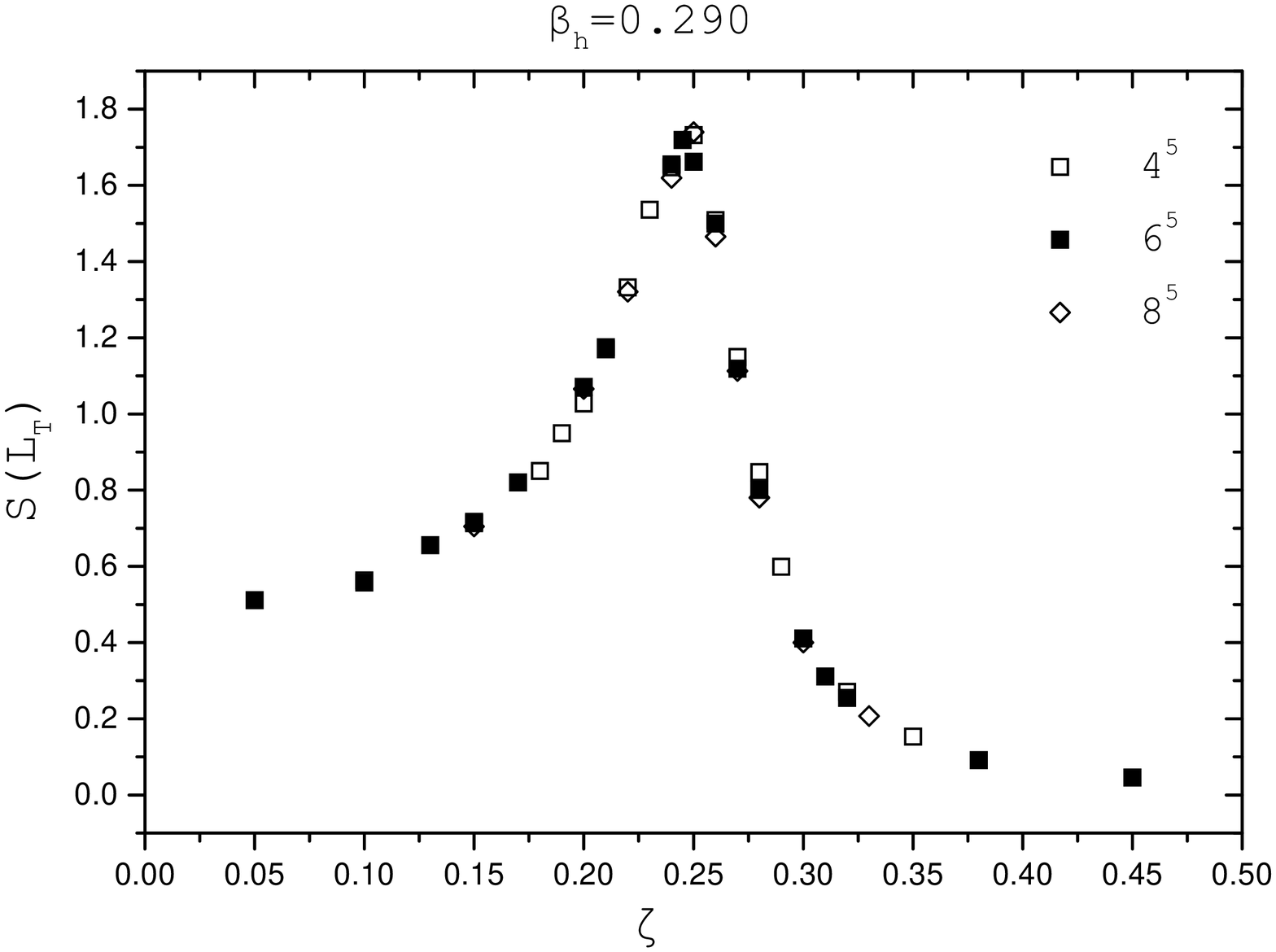,height=14cm,angle=0}}}
\caption[fr4]{Susceptibility of the transverse-like link for $\bt_h=0.29$
and $\beta=4.0.$
\label{susclt290.eps}}
\end{figure}
%%%%%%%%%%%%%%%%%%%%%%%%%%%%%%%%%%%%%%%%%%%%%%%%%%%%%%%%%%%%%%%%

Next we examine the possibility for a non-trivial phase transition
separating the $H_4$ and $H_5$ phases. 
In figure \ref{susclt290.eps} we show the susceptibility of the 
transverse-like link $L_T$ at the $H_4$-$H_5$ transition for 
three lattice volumes,
namely $4^5,~6^5$ and $8^5.$ The data points lie on the same curve; in
particular the peak is the same for the three volumes. Thus 
the data up to the volume $8^5$ suggest that this transition is a crossover.

%%%%%%%%%%%%%%%%%%%%%%%%%%%%%%%%%%%%%%%%%%%%%%%%%%%%%%%%%%%%%%%%
\begin{figure}
\centerline{\hbox{\psfig{figure=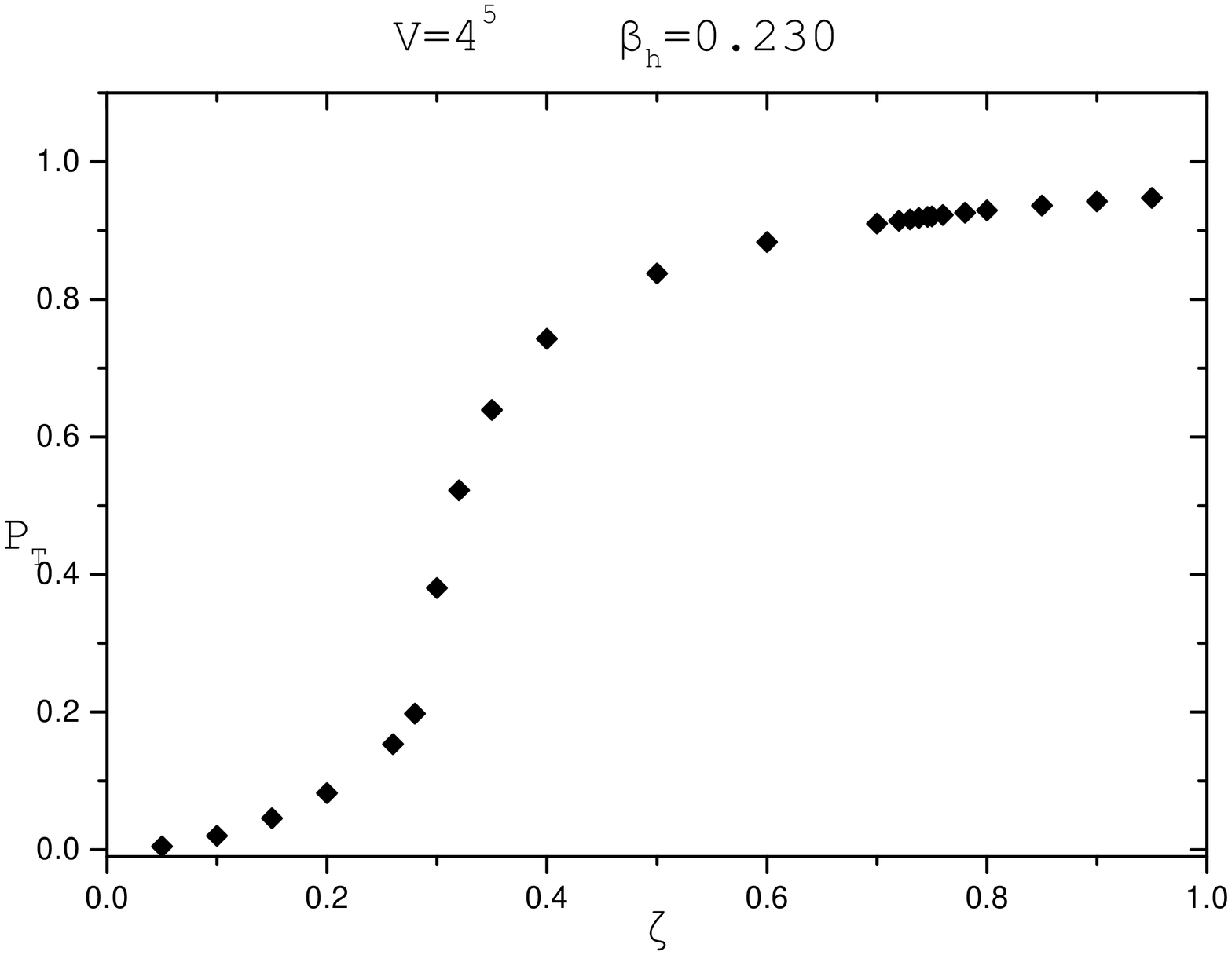,height=10cm,angle=0}}}
\caption[fn4]{Transverse-like plaquette versus $\z$ for a $4^5$
lattice with $\beta=4.0,~~\bt_h=0.23.$}
\label{mptbh0230.eps}
\end{figure}
%%%%%%%%%%%%%%%%%%%%%%%%%%%%%%%%%%%%%%%%%%%%%%%%%%%%%%%%%%%%%%%%

\begin{figure}
\centerline{\hbox{\psfig{figure=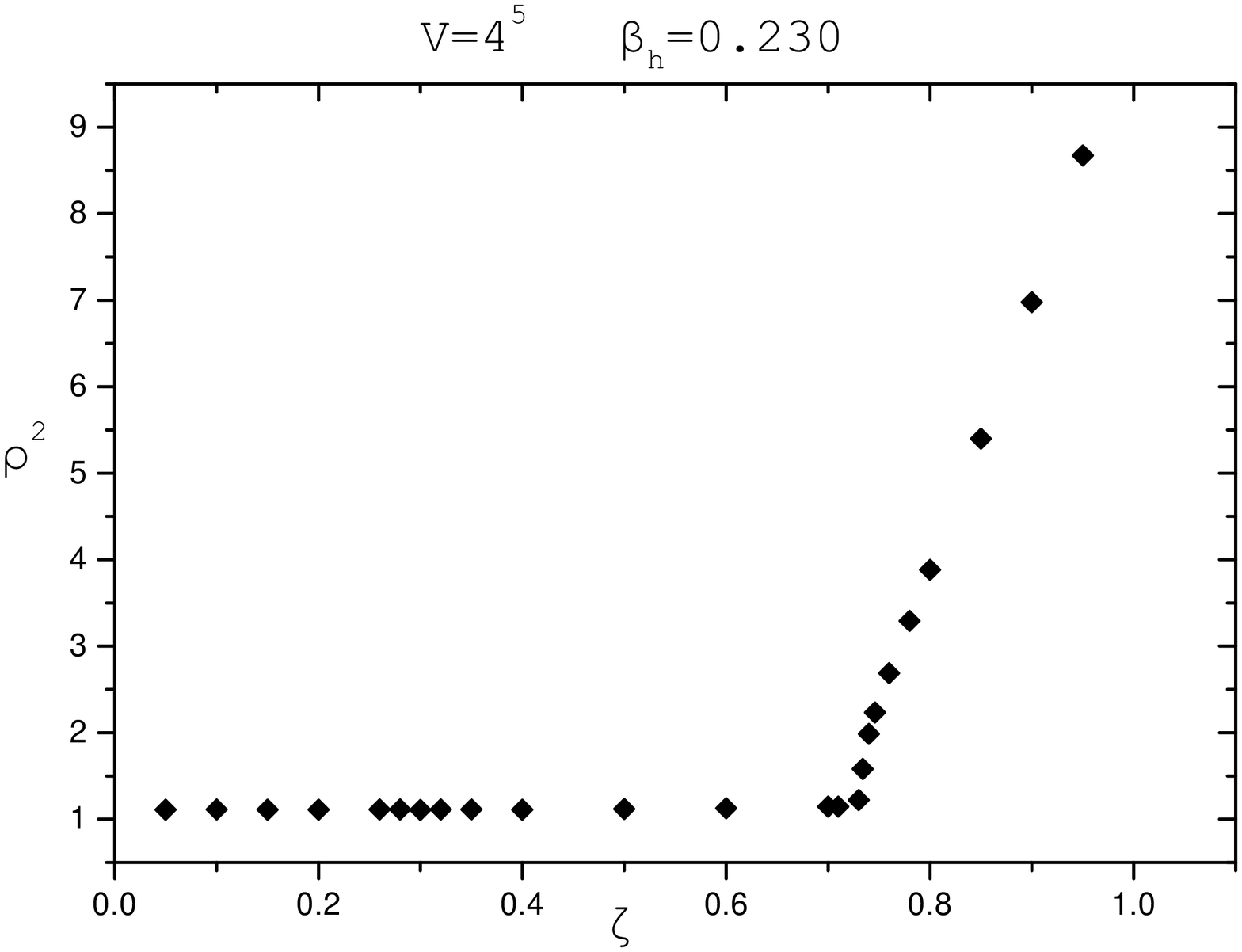,height=10cm,angle=0}}}
\caption[fn5]{$\rho^2$ versus $\z$ for a $4^5$ 
lattice with $\beta=4.0,~~\bt_h=0.23.$}
\label{mr2bh0230.eps}
\end{figure}

The case where $\beta_{h}=0.23$ is more complicated. 
The relevant material is presented in figure \ref{mptbh0230.eps},
where the variation of the transverse-like plaquette
$P_T$ versus $\z$ is shown and figure \ref{mr2bh0230.eps}, which
contains the variation of $\rho^2.$
There are two phase transitions taking place as $\z$ increases. 
The first one takes place at $\z \simeq 0.3,$ as one may see in figure 
\ref{mptbh0230.eps}. 
$P_T$ is tiny for small enough $\z,$ corresponding to
large anisotropy; this indicates that the layers are decoupled for these
values of $\z;$ on the other hand the values of $\rho^2$ turn out to
be relatively small, characterizing the phase as Coulomb. Thus we are in the
four-dimensional Coulomb phase, $C_4$. After the phase transition, the 
value of $P_T$ is sizable, so the layers communicate with each other.
On the other hand, one may see in figure \ref{mr2bh0230.eps} that
$\rho^2$ is still small at this value of $\z.$ 
The result is that the system is in a five-dimensional Coulomb 
phase ($C_5$) and the transition at $\z \simeq 0.3$ separates $C_4$ from $C_5$.

%%%%%%%%%%%%%%%%%%%%%%%%%%%%%%%%%%%%%%%%%%%%%%%%%%%%%%%%%%%%%%%%
%%%%%%%%%%%%%%%%%%%%%%%%%%%%%%%%%%%%%%%%%%%%%%%%%%%%%%%%%%%%%%%%

If we consider even bigger values of $\z$ we find out (as we show in
figure \ref{mr2bh0230.eps}) that $\rho^2$ jumps
to big values at $\z \simeq 0.74.$ Thus, we see the transition
from the five-dimensional Coulomb phase ($C_5$) to the five-dimensional
Higgs phase ($H_5$). Summarizing,  
the system moves first from $C_4$ to $C_5$ and then from $C_5$ to $H_5$, in
full agreement with the mean field predictions.

%%%%%%%%%%%%%%%%%%%%%%%%%%%%%%%%%%%%%%%%%%%%%%%%%%%%%%%%%%%%%%%%
\begin{figure}
\centerline{\hbox{\psfig{figure=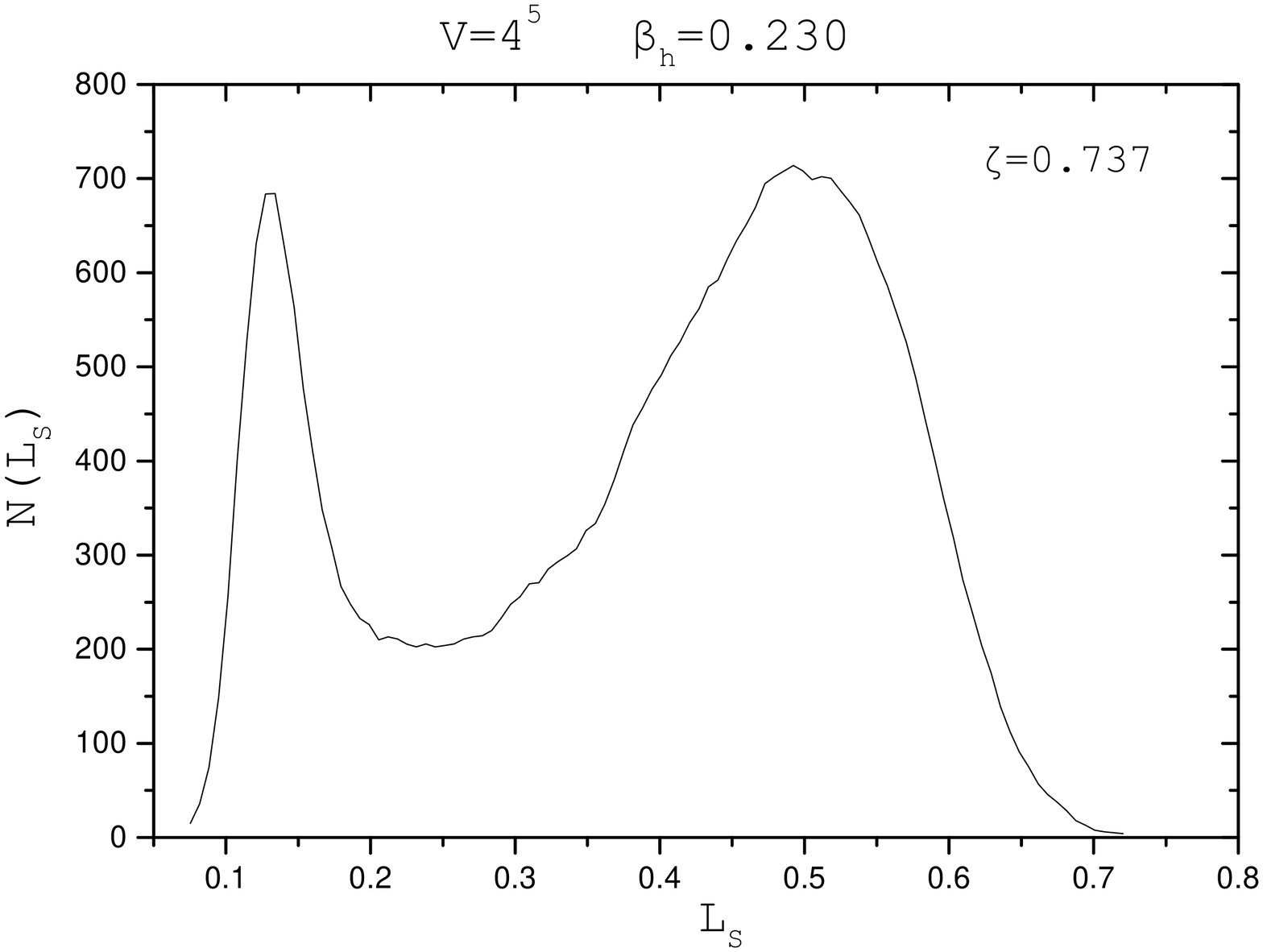,height=10cm,angle=0}}}
\caption[fn6]{Distribution of the space-like link for a $4^5$
lattice at the $C_5$-$H_5$ phase transition.}
\label{distrlsbh0230.eps}
\end{figure}
%%%%%%%%%%%%%%%%%%%%%%%%%%%%%%%%%%%%%%%%%%%%%%%%%%%%%%%%%%%%%%%%

The transition from $C_5$ to $H_5$ is first order. This may be seen from the
distributions of the various observables. In figure 
\ref{distrlsbh0230.eps}
we show the distribution of the space-like link $L_S.$ A clear 
two-peak signal is seen at $\z=0.737,$ so we confirm a strong phase 
transition separating these two phases.

\section{Conclusions}

We have explored the complicated phase structure  
for the Abelian Higgs model in five dimensions allowing for 
anisotropic couplings in the kinetic terms of the Lagrangian. 
This study has been done for a typical  value of the 
(space-like) gauge coupling $\bt$  in the weak coupling regime.
We search for stable four-dimensional layers (3-branes) 
embedded in a five-dimensional world.
Besides the well known 4-D Coulomb phase $C_4$ (where the gauge 
theory is in the symmetric phase with 
a massless photon localized on the 3-brane)   
we have found indications for a new layered phase, denoted by $H_4,$
where the gauge symmetry is broken and the 4-D properties predominate.
The gauge theory is in the confining phase in the bulk. 
For the range of the quartic couplings that we used,  
$C_4$ and $H_4$ are separated by a first order transition.
The $H_4$ and $H_5$ (four- and five-dimensional Higgs) phases 
are separated by a crossover up to the lattice volume we have studied, 
so they appear to be analytically connected;  the realization of 
a continuum four-dimensional  world within a five-dimensional continuum
strongly depends on the parameters of the theory. 
The Monte Carlo results confirm 
qualitatively the predictions we got from the mean field
calculations; the precise characterization of the phase transitions
comes exclusively from the Monte Carlo approach.   

{\bf Acknowledgements}: P. D., K. F. and G. K. acknowledge support from the 
TMR project ``Finite temperature Phase Transitions in Particle Physics'', 
EU contract FMRX-CT97-0122.  
S. N. acknowledges the warm hospitality of the
High Energy Theory Group at the National Technical University of Athens.

\end{document}